\documentclass[useAMS,usenatbib]{mn2e}

\usepackage{graphicx}	
\usepackage{amsmath}	
\usepackage{amssymb}	
\usepackage{multicol}        
\usepackage{bm}		
\usepackage{pdflscape}	
\usepackage{multirow,multicol} 

\def\N1316{NGC\,1316}
\def\N1404{NGC\,1404}

\def\4U{4U~1735$-$444}

\def\aap{{A\&A}}

\def\apjl{{ApJ}}

\def\apjs{{ApJS}}

\def\arcsec{\ifmmode '' \else $''$\fi}

\def\arcsecpoint{\ifmmode ''\!. \else $''\!.$\fi}

\def\kms{\ifmmode {\rm km\ s}^{-1} \else km s$^{-1}$\fi}
\def\Msun{\ifmmode {\rm M}_{\odot} \else M$_{\odot}$\fi}
\def\Lsun{\ifmmode {\rm L}_{\odot} \else L$_{\odot}$\fi}
\def\Zsun{\ifmmode {\rm Z}_{\odot} \else Z$_{\odot}$\fi}

\def\ergscm2{ergs\,s$^{-1}$\,cm$^{-2}$}
\def\icm3{{\rm cm}^{-3}}
\def\icm2{{\rm cm}^{-2}}
\def\qo{\ifmmode q_{\rm o} \else $q_{\rm o}$\fi}
\def\Ho{\ifmmode H_{\rm o} \else $H_{\rm o}$\fi}
\def\ho{\ifmmode h_{\rm o} \else $h_{\rm o}$\fi}

\def\vFWHM{\ifmmode v_{\mbox{\tiny FWHM}} \else
            $v_{\mbox{\tiny FWHM}}$\fi}
\def\CCF{\ifmmode F_{\it CCF} \else $F_{\it CCF}$\fi}
\def\ACF{\ifmmode F_{\it ACF} \else $F_{\it ACF}$\fi}
\def\Halpha{\ifmmode {\rm H}\alpha \else H$\alpha$\fi}
\def\Hbeta{\ifmmode {\rm H}\beta \else H$\beta$\fi}
\def\Hgamma{\ifmmode {\rm H}\gamma \else H$\gamma$\fi}
\def\Hdelta{\ifmmode {\rm H}\delta \else H$\delta$\fi}
\def\Lya{\ifmmode {\rm Ly}\alpha \else Ly$\alpha$\fi}
\def\Lyb{\ifmmode {\rm Ly}\beta \else Ly$\beta$\fi}
\def\Lyg{\ifmmode {\rm Ly}\beta \else Ly$\gamma$\fi}

\def\heii{He\,{\sc ii}}

\def\ciii{\ifmmode {\rm C}\,{\sc iii} \else C\,{\sc iii}\fi}
\def\civ{\ifmmode {\rm C}\,{\sc iv} \else C\,{\sc iv}\fi}
\def\cv{\ifmmode {\rm C}\,{\sc v} \else C\,{\sc v}\fi}
\def\cvi{\ifmmode {\rm C}\,{\sc vi} \else C\,{\sc vi}\fi}

\def\oiii{O\,{\sc iii}}
\def\o5007{[O\,{\sc iii}]\,$\lambda5007$}

\def\ovi{O\,{\sc vi}}

\def\oviii{O\,{\sc viii}}

\def\nex{Ne\,{\sc x}}

\def\fexxii-iii{Fe\,{\sc xxii-xxiii}}

\usepackage{hyperref}
\hypersetup{colorlinks=true,linkcolor=blue,citecolor=blue,filecolor=blue,urlcolor=blue}

\title[AGN feedback in the Phoenix cluster]{AGN feedback in the Phoenix cluster}
\author[C. Pinto et al.]{C. Pinto,$^{1}$\thanks{E-mail:
cpinto@ast.cam.ac.uk} C. J. Bambic,$^{1,2}$ J. S. Sanders,$^{3}$ A. C. Fabian,\,$^{1}$ \newauthor
M. McDonald,$^{4}$ H. R. Russell,$^{1}$ H. Liu,$^{1}$ and C. S. Reynolds,$^{1,2}$\\
$^{1}$Institute of Astronomy, Madingley Road, CB3 0HA Cambridge, United Kingdom\\
$^{2}$Department of Astronomy, University of Maryland, College Park, MD 20742-2421, USA\\
$^{3}$Max-Planck-Institut f\"ur extraterrestrische  Physik, Giessenbachstrasse 1, 85748 Garching, Germany\\
$^{4}$Kavli Institute for Astrophysics and Space Research, MIT, 77 Massachusetts Avenue, Cambridge, MA 02139, USA}
\begin{document}

\date{Accepted 2018 August 8. Received 2018 August 3; in original form 2018 April 20}

\pagerange{\pageref{firstpage}--\pageref{lastpage}} \pubyear{2018}

\maketitle

\label{firstpage}

\begin{abstract}
Active galactic nuclei (AGN) release a huge amount of energy into the intracluster medium (ICM)
with the consequence of offsetting cooling
and star formation (AGN feedback) in the centers of cool core clusters. 
The Phoenix cluster is among the most massive clusters of galaxies known in the 
Universe. It hosts a powerful starburst of several hundreds of Solar masses per year 
and a large amount of molecular gas in the center. 
In this work we use the high-resolution Reflection Grating Spectrometer (RGS) on board 
XMM-\textit{Newton} to study the X-ray emitting cool gas in the Phoenix cluster
and heating-cooling balance. 
We detect for the first time evidence of {O\,\scriptsize{VIII}} and 
{Fe\,\scriptsize{XXI-XXII}} emission lines, 
{the latter} demonstrating the presence of gas below 2 keV.
We find a cooling rate of 
$350\pm_{200}^{250}\,M_{\odot}\,{\rm yr}^{-1}$ below 2\,keV (at the 90\% confidence level), 
which is consistent with the star formation rate in this object.
This cooling rate is high enough to produce the molecular gas found in the filaments 
via instabilities during the buoyant rising time.
The line broadening indicates that the turbulence ($\sim300$\,km\,s$^{-1}$ or less) is below 
the level required to produce and propagate the heat throughout the cool core.
This provides a natural explanation for the coexistence of large amounts of cool gas,
star formation and a powerful AGN in the core.
The AGN activity may be either at a young stage
or in a different feedback mode, due to a high accretion rate.
\end{abstract}


\section{Introduction}
\label{sec:intro}

Clusters of galaxies are extraordinary laboratories where highly energetic astrophysical
phenomena are in action. The study of the intracluster medium (ICM) embedded in their 
deep gravitational well allow us to understand the role of the strong energetic throughput
from the active galactic nuclei (AGN) present in the individual galaxies into their surroundings over
large scales up to a hundred kpc or more. A substantial fraction of galaxy clusters show
a highly peaked density profile which implies that the central cooling time can be of an order
of magnitude lower than the current age of the Universe (cool core clusters, 
e.g. \citealt{Hudson2010}).
In the absence of heating, this would imply the cooling of hundreds of Solar masses 
of gas per year below $10^{6}$\,K \citep{Fabian1994} for the massive clusters,
with a consequent star formation rate of a similar order of magnitude.

The best UV and X-ray indicators of cool gas are O\,{\scriptsize VI} emission lines 
(peaking at $T\sim3\times10^{5}$\,K), O\,{\scriptsize VII} ($T\sim2\times10^{6}$\,K) 
and Fe\,{\scriptsize XVII} ($T\sim6\times10^{6}$\,K).
High-resolution UV and X-ray spectra of ICM were therefore expected to show
bright lines, which instead turned out to be much weaker
at levels of about $30\,M_{\odot}$\,yr$^{-1}$ or lower 
(see e.g. \citealt{Bregman2005,Bregman2006}, 
\citealt{Peterson2003} and \citealt{Pinto2014}). 

There is a wealth of highly energetic phenomena occurring 
in the cores and in the outskirts of clusters of galaxies.
Galactic mergers and gas sloshing within the gravitational potential 
may release large amounts of energy in the outskirts via shocks and turbulence injection
(see e.g. \citealt{Ascasibar2006}; \citealt{Lau2009}).
However, there is a growing consensus that AGN are the most
relevant sources of heating in cluster cores, particularly through their powerful relativistic jets 
(AGN kinetic / radio mode feedback, see e.g.
\citealt{Churazov2000} and \citealt{Fabian2012}). 
For instance, the work done by AGN to inflate bubbles in the surrounding ICM gas proves 
to be equal if not stronger than the cooling rates \cite{McNamara2007}.
There are several mechanisms through which the energy can be released into the ICM. 
Two elegant solutions involve dissipation of turbulence 
\citep[see e.g.][]{Zhuravleva2014} or sound waves
(see e.g.  \citealt{Fabian2003_waves, Fabian2017sw}).

There seems to be evidence that the extreme radio mode of AGN feedback has been 
operating in a steady way for the past 5 Gyr (see, e.g., \citealt{Hlavacek-Larrondo2012} 
and \citealt{McDonald2013}), but this research field is still rather young and in development.
Measurements of AGN-induced turbulence are therefore key to test whether some 
scenarios of AGN feedback and cooling-heating balance are feasible.
This is crucial to understand if there is enough energy in the ICM
to propagate the heat throughout the cool core.

It is possible to place constraints on turbulence by measuring the velocity broadening 
of the X-ray emission lines produced by the hot ICM. 
The Reflection Grating Spectrometers (RGS, \citealt{denherder2001}) on board 
XMM-\textit{Newton} are currently the only X-ray instruments which have enough 
collecting area and spectral resolution to enable this measurement.
However, this is not very straightforward because the spectrometers do not possess 
an appropriate slit and therefore instrumental broadening has to be accounted for.
\citet{Sanders2010} made the first measurement of cluster velocity broadening
using the luminous cluster A\,1835 at redshift 0.25. Due to the limited spatial extent 
of its bright core, instrumental broadening was minimal and 
an upper limit of 274\,km\,s$^{-1}$ was obtained.
\citet{Sanders2011} constrained turbulent velocities for a large sample
of 62 clusters, groups, and elliptical galaxies observed with XMM-\textit{Newton}/RGS.
Half of them show velocity broadening below 700\,km\,s$^{-1}$. Recently, 
\citet{Sanders2013} used continuum-subtracted emission line surface brightness
profiles to account for the spatial broadening. 
\citet{Pinto2015} focused on nearby objects using a catalog of 44 sources, 
the CHEERS sample, consisting of bright clusters, groups of galaxies, elliptical galaxies
with a $\gtrsim5\sigma$ detection of the O\,{\scriptsize VIII} 1s--2p line at 19\,{\AA}
and with a well-represented variety of strong and weak cool-core objects. 
They confirmed the results obtained by \citet{Sanders2013} in more distant objects
despite the more severe instrumental broadening due to the short distances ($z<0.1$).
\citet{Pinto2015} also showed that the upper limits on the Mach numbers are typically larger
than the values required to balance cooling, suggesting that dissipation of turbulence may 
be high enough to heat the gas and to prevent cooling.

Turbulence in giant elliptical galaxies has also been constrained with an alternative method
which uses the ratio of the Fe\,{\scriptsize{XVII}} emission lines detected in the RGS spectra
(see, e.g., \citealt{Werner2009}, \citealt{dePlaa2012}, \citealt{Pinto2016mnras}, \citealt{Ogorzalek2017}).
When the velocity broadening is low, the gas is optically thick in the 15\,{\AA} line 
due to resonant scattering, while the 17\,{\AA} lines remain optically thin. 
Comparison between the observed line ratios with simulations for different Mach numbers 
constrains the level of turbulence.
This method is very efficient for cool ($kT<0.9$\,keV) giant elliptical galaxies rich in Fe\,{\scriptsize{XVII}}
emission lines, but it is significantly limited by the systematic uncertainty ($\sim$20\%) 
in the line ratio for an optically thin plasma. Currently, there are no facilities that allow 
the use of this technique on clusters of galaxies since the optical depth of the higher ionisation lines,
typical of clusters, are smaller than those of the Fe\,{\scriptsize{XVII}} lines or they are out of the 
RGS energy band. 

Another alternative method to constrain turbulence interprets the surface brightness fluctuations 
commonly seen in X-ray atmospheres of clusters as turbulent fluctuations (see, e.g.,
\citealt{SandersFabian2012}, \citealt{Zhuravleva2014}, \citealt{Walker2015} and \citealt{Eckert2017}). 
The measurements imply motions of one to a few hundred km\,s$^{-1}$.

As of today, and likely for the next years until the launch of missions like 
\textit{XRISM} (a.k.a. XARM) and \textit{ATHENA} 
(for a review see, e.g., \citealt{Nandra2013} and \citealt{Guainazzi2018}), 
the most accurate measurement of line broadening and, therefore,
constraint on turbulence, has been obtained by the {\textit{Hitomi}} observations of the Perseus
cluster of galaxies \citep{Hitomi2016nat}. The soft X-ray microcalorimeter  
(SXS) onboard {\textit{Hitomi}}, the first successfully operative in space,  
measured an average line broadening of $164\pm10$\,km\,s$^{-1}$ in a region between
30 and 60 kpc from the central AGN,
thanks to an astonishing and unprecedented high spectral resolution of 4.9\,eV
 in the Fe K energy band.
Further work accounting for PSF effects showed that in several regions of the cluster core
the turbulence could be lower than 100 km\,s$^{-1}$ \citep{Hitomi2017atm}.
These motions may not be able to propagate throughout the cluster rapidly enough to offset 
radiative losses at each radius of the cluster \citep{Fabian2017sw}.

In parallel work (Pinto et al. in prep) we have made a large catalog using all the 
XMM-\textit{Newton} observations of clusters and groups of galaxies and ellipticals. 
The main goal is to constrain 
turbulence and other physical characteristics of a statistical sample of 150 galaxy clusters 
up to redshift 0.6 using high-resolution RGS spectra.
This catalog includes and significantly extends what was previously done in \cite{Sanders2013}
and \citet{Pinto2015}. 
We also use a new technique to account for instrumental broadening 
that has been introduced in a very recent paper \citep{Bambic2018}.  
It was tested on three clusters at different redshifts with an
observation quality among the best in our sample 
(A\,1835, A\,2204, and MACS\,J2229.7-2755).
Upper limits of 200--250\,km\,s$^{-1}$ were obtained on turbulence, 
confirming that it might be not high
enough to propagate the heat throughout the cool core.

In this work we study the Phoenix cluster (SPT-CLJ2344-4243), 
the most distant object in our XMM-\textit{Newton}/RGS catalog ($z=0.596$), 
the most luminous X-ray cluster known and one of the most massive 
clusters ($\sim 2\times10^{15}\,M_{\odot}$, see \citealt{Williamson2011} and
\citealt{McDonald2012}).
It also exhibits a starburst of $500-800\,M_{\odot}\,{\rm yr}^{-1}$ 
(see, e.g., \citealt{McDonald2015} and \citealt{Mittal2017}),
which is among the largest found in the local Universe ($z<1$).
\citet{McDonald2015} found large amounts of mildly-ionized gas,
such as {\heii}, {\oiii} and {\ovi}, which is not consistent with cooling 
but rather suggests that an ionized wind is driven 
by the AGN and/or the starburst.
ALMA observations of the CO(3-2) line emission have shown a huge amount
of molecular gas around $2\times10^{10}\,M_{\odot}$ (\citealt{Russell2017}).
The molecular gas might have been uplifted directly by the radio bubbles
or formed via instabilities in low entropy gas lifted by the jet.
\citet{Tozzi2015} observed the Phoenix cluster for 250 ks with XMM-\textit{Newton}
and found a cooling rate between 100--1000\,$M_{\odot} {\rm yr}^{-1}$ 
with the large uncertainties mainly driven by the calibration between the CCD cameras 
on board the satellite combined with their low spectral resolution.
They also used the RGS gratings and claimed to have not found any lines
from ionization states below {Fe\,\scriptsize{XXIII}} with an upper limit 
on the cooling rate of about $500\,M_{\odot}\,{\rm yr}^{-1}$ below 3\,keV.

The presence of a high star formation, a high cooling rate and a powerful AGN
in the Phoenix cluster casts doubts on the efficiency of AGN feedback 
to balance cooling. 
Therefore, we perform an in-depth study of the XMM-\textit{Newton}
observations of the Phoenix cluster with particular focus on the high-resolution
spectrometers in order to obtain more accurate measurements of the cooling rate
and the turbulent broadening.
We find that $350\pm_{120}^{150} (68\%) \pm_{200}^{250} (90\%)\,M_{\odot}\,{\rm yr}^{-1}$ are cooling below 2\,keV 
in the cluster core and that the turbulence level is likely not adequate 
to fully propagate AGN heating throughout the cool core.
The cooling rate is instead high enough to produce the large amount of 
molecular gas found in the form of filaments if we assume that the gas cools 
during the bubble rising time.

We present the data in Sect.\,\ref{sec:data} and the spectral modeling in  
Sect.\,\ref{sec:spectral_modeling}. We discuss the results in Sect.\,\ref{sec:discussion} 
and give our conclusions in Sect.\,\ref{sec:conclusion}.

\begin{table}
\caption{XMM-\textit{Newton} and \textit{Chandra} log of observations.}  
\label{table:log}      
\renewcommand{\arraystretch}{1.1}
 \small\addtolength{\tabcolsep}{-4pt}
 \centering
\scalebox{1}{%
\begin{tabular}{cccc|ccc}     
\hline  
OBS\_ID        & t$_{\rm RGS}$\,$^{(a)}$ & c$_{\rm RGS}$\,$^{(c)}$ & t$_{\rm MOS}$\,$^{(b)}$ &  &  OBS\_ID     & t$_{\rm ACIS}$\,$^{(d)}$ \\  
 (XMM)          & (ks)                                 & (counts)                           &  (ks)            &        &  \textit{Chandra}  & (ks)  \\ 
\hline                                                                                
0693661801  &     16.5                            &       1.8k                            &   16.1   &  &   13401  & 11.9   \\   
0722700101  &   130.6                            &     14.1k                            & 128.0   &  &   16135  & 57.3   \\   
0722700201  &     93.0                            &     10.1k                            &   92.0   &   &  16545  & 59.1   \\   
\hline                
\end{tabular}}

$^{(a)}$ RGS1, RGS2 and $^{(b)}$ MOS1 net exposure time.  \\ 
$^{(c)}$ RGS total counts and $^{(d)}$ ACIS-I net exposure time.
\end{table}
 
\section[]{The data}
\label{sec:data}

The observations used in this paper are listed in Table~\ref{table:log}. 
The XMM-\textit{Newton} satellite is provided with two main X-ray instruments: 
RGS (Reflection Grating Spectrometer) and EPIC (European Photon Imaging Camera).
We use RGS for the spectral analysis and EPIC (MOS\,1 detector, 
which is aligned with RGS) for imaging.
The EPIC/MOS\,1 surface brightness profiles are necessary to account for 
line instrumental broadening due to the slitless nature of the RGS spectrometers.
This is done following the procedures used in \citet{Pinto2015} with important
updates shown in \citet{Bambic2018}. 

We perform the data reduction with the XMM-\textit{Newton} Science Analysis System 
({\scriptsize{SAS}}) v16, CalDB as of January 2018.
We process the RGS data with the SAS task \textit{rgsproc} 
and the MOS\,1 data with \textit{emproc} to produce event files, 
spectra, and response matrices for RGS (both 1 and 2) and images for MOS\,1.
Following the standard procedures, 
we filter the MOS event list for bad pixels, bad columns, cosmic-ray events 
outside the field of view (FOV), photons in the gaps (FLAG = 0), 
and apply standard grade selection, corresponding to PATTERN $\leq12$.

We correct for contamination from soft-proton flares through the SAS task 
\textit{evselect} by extracting light curves for MOS\,1 in the 10--12 keV
energy band, while we use the data from the CCD number 9 for RGS,
where hardly any emission from the source is expected. 
The light curves are grouped in 100\,s intervals and all the time bins 
with a count rate above 0.35 c/s and 0.15 c/s are rejected for MOS and RGS, respectively. 
We build the good time interval (GTI) files with the accepted time 
events for the MOS and RGS data through the SAS task \textit{tabgtigen} and 
reprocess the data again with \textit{rgsproc} and \textit{emproc}. 
For RGS 1 and 2 we join the GTI and obtain the same exposure times.
The RGS\,1-2 and MOS\,1 total clean exposure times are quoted in Table\,\ref{table:log}.

\subsection[]{RGS spectra}
\label{sec:data_spectra}

The RGS 1 and 2 spectra are extracted using as centroid 
$(\alpha, \delta)=(23:44:43.9,-42:43:13.7)$ 
and a width of 50 arcsec 
by adopting the mask \textit{xpsfincl} $=90$ in the \textit{rgsproc} 
($\sim\pm170$\,kpc for a standard $\Lambda$CDM cosmology with 
$H_{\rm 0} = 70\,$km\,s$^{-1}$ Mpc$^{-1}$, $\Omega_{\rm M} = 0.27$ 
and $\Omega_{\rm \Lambda} = 0.73$).
In order to subtract the background we test both the standard background
spectrum which is extracted beyond the 98\% of the RGS point spread function 
(PSF, \textit{xpsfexcl} $=98$ in the \textit{rgsproc}) and the model background spectrum, 
which is a template background file based on the count rate in CCD\,9.
Normally one would use the model background for an extended source, 
but due to the large distance we can safely use the observation background 
since it is extracted beyond $\pm1.4'$.
The background spectra are indeed comparable, provide consistent results
and about 26,000 net source counts in the RGS 6--33\,{\AA} wavelength band.
{More detail on the background is provided in Appendix\,\ref{sec:appendix}.}

\begin{figure}
  \includegraphics[width=1\columnwidth, angle=0]{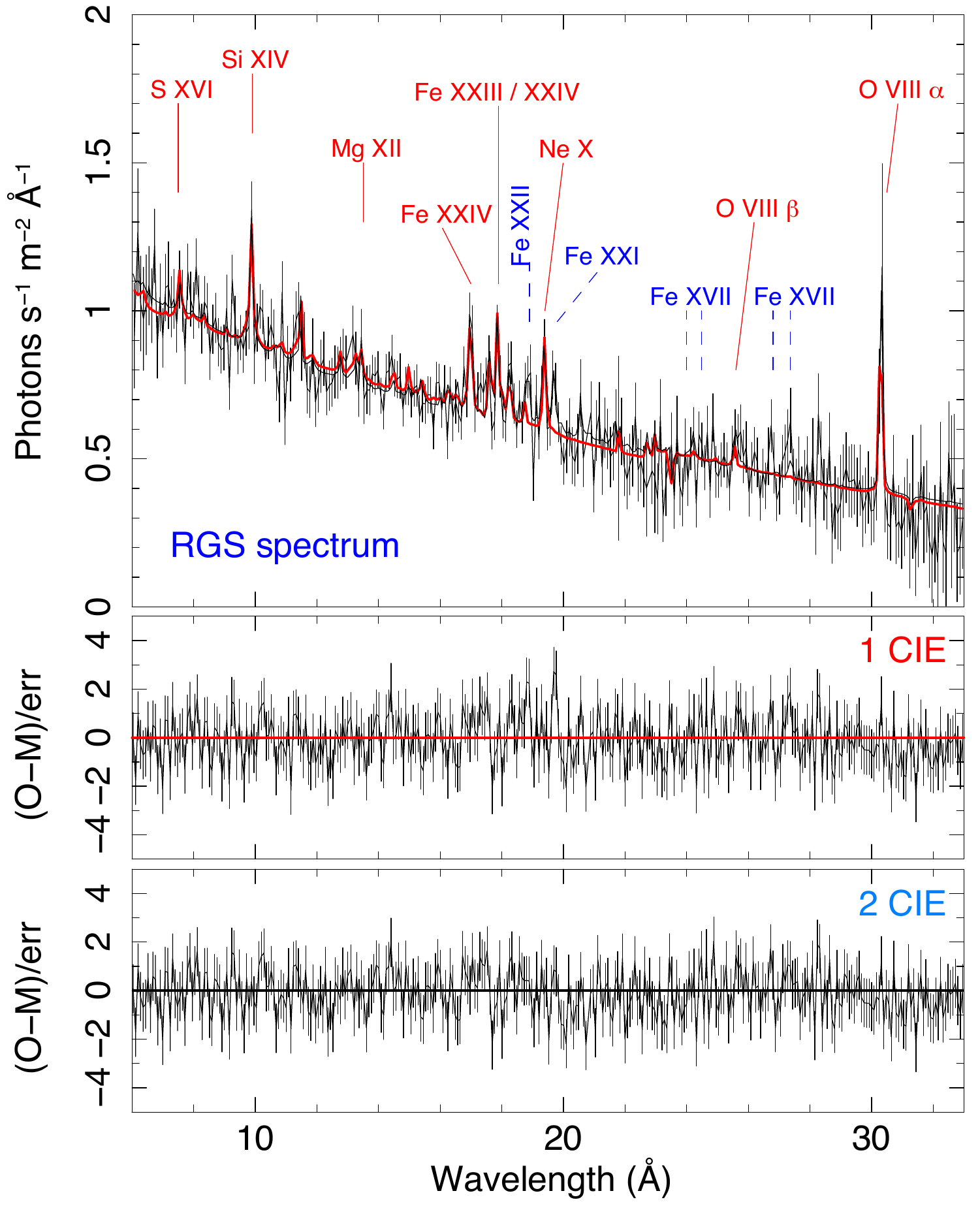}
   \caption{RGS spectrum of the Phoenix cluster with overlaid an isothermal model of gas
   in collisional equilibrium (red) and a two-temperature model (black). 
   Emission lines commonly detected in RGS spectra of cool core 
   clusters are labelled at the observed wavelengths
   {with the blue dashed ones referring to lines from cool gas.
   Bottom plots show the residuals to each model. 
   Note how the second component improves the fit near the {Fe\,\scriptsize{XVII}} and 
   {Fe\,\scriptsize{XXI-XXII}} main transitions.}} 
   \label{Fig:Phoenix_RGS_spectrum}
\end{figure}

The spectra are converted to the {\scriptsize SPEX}\,\footnote{www.sron.nl/spex} 
format through the {\scriptsize SPEX} task \textit{trafo}.
During the spectral conversion, we choose the option of \textit{sectors} 
in the task \textit{trafo} to create as many sectors as there are different exposures. 
This permits us to simultaneously fit the multiple RGS spectra 
by choosing which parameters to either couple or unbind 
in the spectral models of different observations.
We focus on the first-order spectra because the second-order spectra have
too poor statistics. We also combine source and background spectra, 
and responses from all observations and from both RGS instruments 
using the SAS task \textit{rgscombine}
for plotting purposes only. 
The stacked spectrum is shown in Fig.\,\ref{Fig:Phoenix_RGS_spectrum} 
labelling the rest-frame wavelengths of the 
strongest emission lines commonly found 
in deep RGS spectra of clusters and groups of galaxies. 

\subsection[]{MOS images}

\begin{figure}
 \centering
  \includegraphics[width=0.9\columnwidth, angle=0]{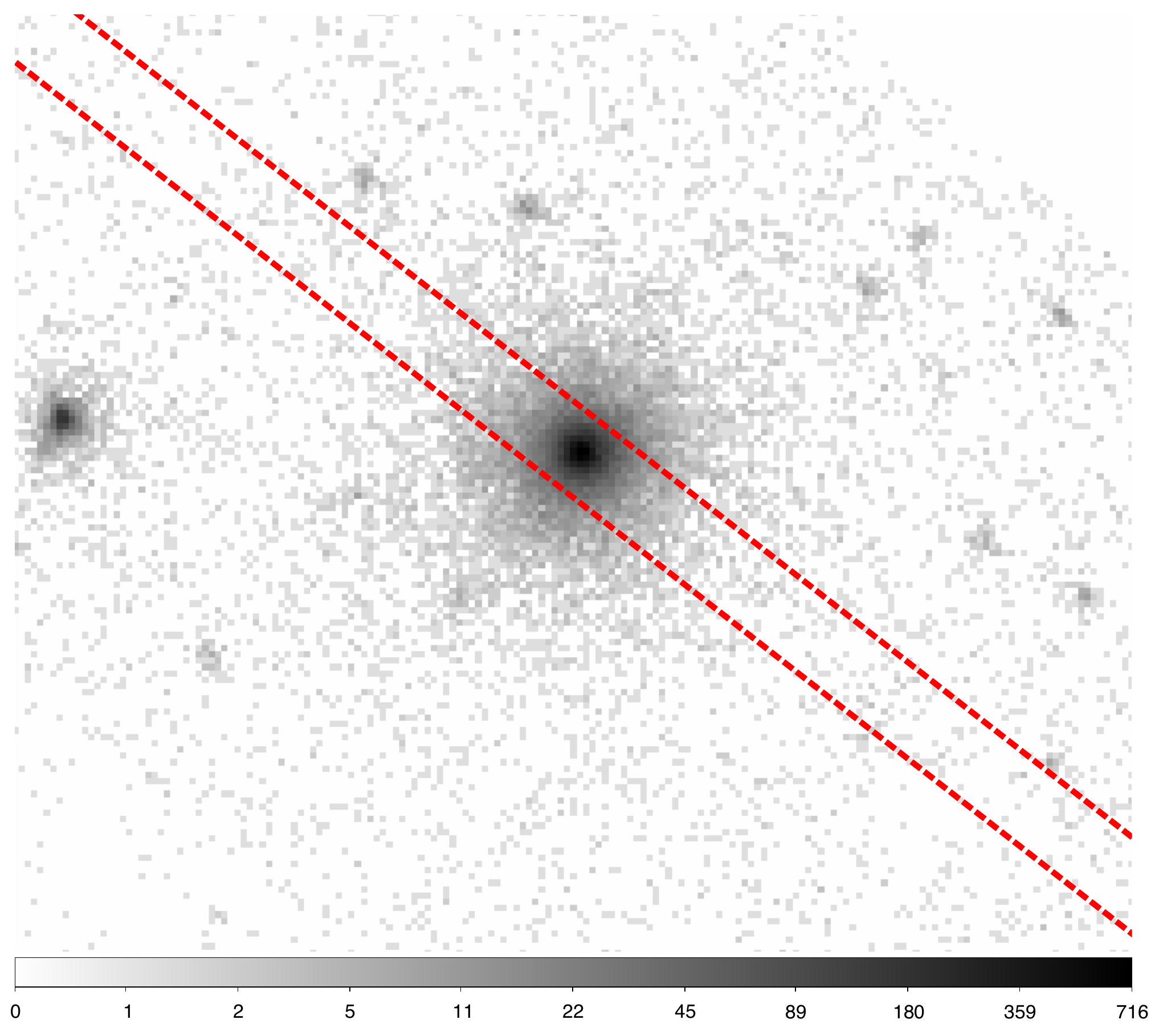}
   \caption{EPIC/MOS\,1 image of the Phoenix cluster with the RGS extraction region 
   of width of 50 arcsec.} \label{Fig:Phoenix_MOS_image}
\end{figure}

\begin{figure}
 \centering
  \includegraphics[width=0.95\columnwidth, angle=0]{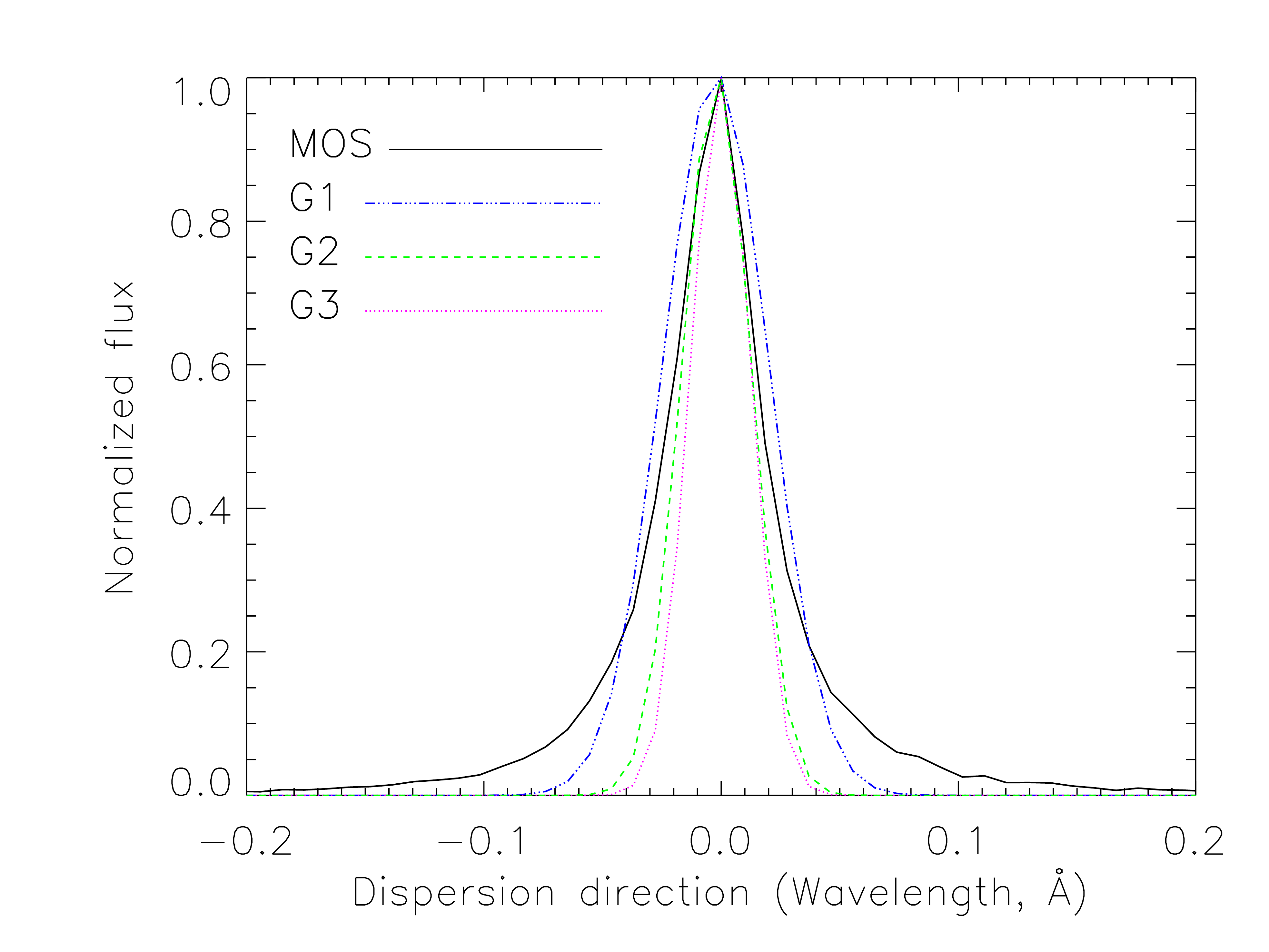}
   \caption{RGS line spatial broadening as computed through the MOS\,1 surface 
   brightness profile and Eq.\,\ref{Eq:disp}. 
   The lines show the narrowest Gaussian obtained by fitting three different models
   with either one or two or three Gaussian lines.} \label{Fig:Phoenix_MOS_profile}
\end{figure}

We produce MOS\,1 images in the $6-35$\,{\AA} wavelength band 
(see Fig.\,\ref{Fig:Phoenix_MOS_image}), which is the same
range used for the RGS spectroscopy, and extract surface brightness profiles 
to model the RGS line spatial broadening with the standard dispersion equation: 
\begin{equation}  \label{Eq:disp}
\Delta\lambda = 0.138 \, \Delta\theta \, {\mbox{\AA}} / m, 
\end{equation}
where $\Delta\lambda$ is the wavelength broadening, 
$\Delta\theta$ is the source extent in arcseconds
and $m$ is the spectral order 
(see the XMM-\textit{Newton} Users Handbook).
The surface brightness is extracted using a region of $50''$ width and
a length of $10'$ with the {\scriptsize SPEX} task \textit{rgsvprof}
(see, e.g., \citealt{Pinto2015}).
The RGS instrumental line broadening as measured through the MOS\,1 surface
brightness profile of the Phoenix cluster is shown in 
Fig.\,\ref{Fig:Phoenix_MOS_profile} (black solid line).

We also fit the broadening profile with 3 different models using 
{\scriptsize QDP/PLT\footnote{https://wwwastro.msfc.nasa.gov/qdp/}}: 
a single Gaussian, two Gaussian and three Gaussian lines. 
The single Gaussian line and the narrowest
Gaussian of the multi component models are also shown 
in Fig.\,\ref{Fig:Phoenix_MOS_profile}, but
renormalized for plotting purposes.

\section{Spectral analysis}
\label{sec:spectral_modeling}

\subsection{Isothermal collisional equilibrium model}
\label{sec:baseline_model}

\begin{table}
\caption{XMM-\textit{Newton}/RGS best-fit isothermal \textit{cie} model.}  
\label{table:rgs_bestfit}      
\renewcommand{\arraystretch}{1.}
 \small\addtolength{\tabcolsep}{-3pt}
\scalebox{1}{%

\begin{tabular}{c c c c c}     
\hline  
         $n_{\rm e} n_{\rm H} V$   & $kT$              & $v_{\rm mic} (2D)$   &  $C/d.o.f.$  & O/H                  \\  
         $4.52\pm0.06$                 & $7.0\pm0.5$  &  $230\pm220$          &   1105/951  & $0.38\pm0.12$  \\
\hline                                               
  Ne/H                 & Mg/H & Si/H & S/H & Fe/H  \\  
  $0.43\pm0.12$ & $0.45\pm0.25$ & $0.75\pm0.16$ & $0.5\pm0.3$ & $0.8\pm0.2$ \\ 
\hline                                               
\end{tabular}}

Abundances are in proto-Solar units \citep{Lodders09},
emission measure $n_{\rm e} n_{\rm H} V$ in $10^{74} {\rm m}^{-3}$, temperature in keV
and micro-turbulence 2D velocity in km s$^{-1}$ as default in {\scriptsize{SPEX}}.
\end{table}


Our analysis focuses on the $6-35$\,{\AA} first order RGS spectra,
where the source counts are above the background and cover the $4-22$\,{\AA}
rest-frame wavelength range. This includes K\,shell emission lines of 
sulphur, silicon, magnesium, neon, oxygen and the L\,shell lines of iron and nickel.
We perform the spectral analysis with {\scriptsize SPEX}
version 3.04.00. We scale the elemental abundances to the proto-Solar abundances 
of \citet{Lodders09}, which are the default in {\scriptsize SPEX}, 
use C-statistics and adopt $1\,\sigma$ errors, 
unless otherwise stated.

We describe the ICM emission with an isothermal plasma model 
of collisional ionization equilibrium (\textit{cie}). 
The basis for this model is given by the {\scriptsize MEKAL} code, 
but several updates have been included (see the {\scriptsize SPEX} manual).
Free parameters in the fits are the emission measure $Y=n_{\rm e}\,n_{\rm H}\,V$, 
the temperature $T$, the micro-turbulence 2D velocity broadening 
and the abundances (O, Ne, Mg, Si, S and Fe).
The abundances of all the other elements are coupled to iron.
The \textit{cie} emission model is corrected for redshift (0.596) and
Galactic absorption ($N_{\rm H}=1.7\times10^{24}$\,m$^{-2}$, \citealt{Kalberla2005})
through the {\textit{hot}} model with very low temperature 
(0.5 eV). This component accounts for both absorption edges and lines of neutral
and low-ionisation species in the ISM (see, e.g., \citealt{Pinto2013}).

We do not explicitly model the cosmic X-ray background in the RGS spectra 
because any diffuse emission feature would be smeared out into a broad continuum-like component. 
The Phoenix cluster is known to host a powerful highly-obscured AGN; 
its emission is negligible in the soft ($<2$\,keV) X-ray band 
{(see \citealt{Tozzi2015} and references therein)} and 
we therefore do not include it in our model.
For each exposure, we simultaneously fit the RGS 1 and 2 spectra
by adopting the same model, apart from the emission measures of the \textit{cie} components
which are uncoupled to account for the slightly different roll angles of the observations.

This simple plasma model provides a reasonably good fit of the RGS spectra
as shown by \citet{Tozzi2015}. Most strong lines are well described by the isothermal model, 
see Fig.\,\ref{Fig:Phoenix_RGS_spectrum} and Table\,\ref{table:rgs_bestfit}. 
However, there are some emission-like features
missed just near the rest-frame wavelengths of the strongest transitions 
of Fe\,{\scriptsize{XXI}} and {\scriptsize{XXII}}. 
It is therefore an interesting and useful exercise to formally search for indicators 
of cool gas before calculating any cooling rates. 
One may in principle just add a few Gaussian lines corresponding to the strongest transitions
expected from gas cooler than that contributing to the strong Fe\,{\scriptsize{XXIII-XXIV}} lines.
 
\subsection{Search for cool gas}
\label{sec:search_cool_gas}

\begin{figure}
  \includegraphics[width=0.95\columnwidth, angle=0]{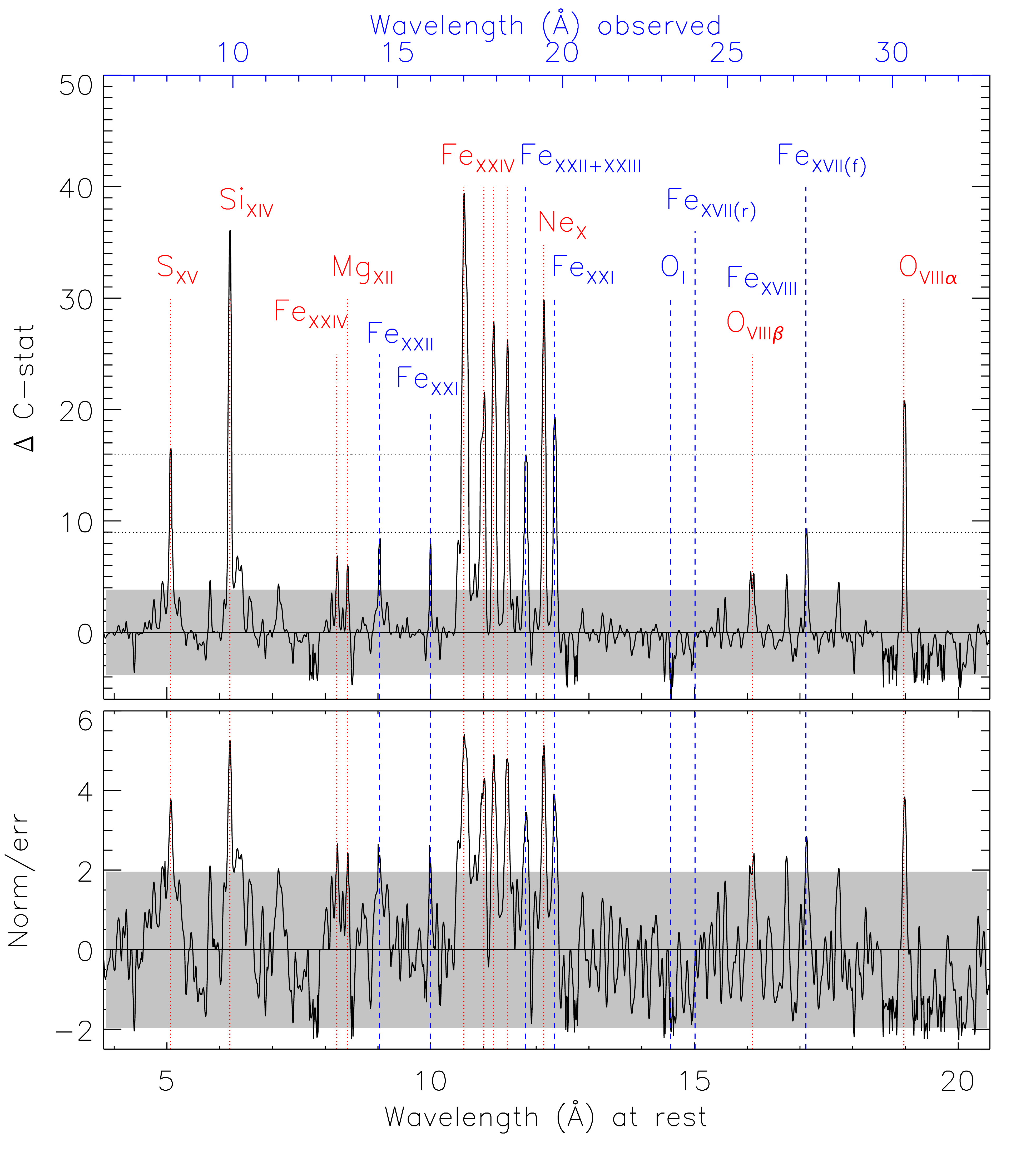}
   \caption{RGS Gaussian line search. 
            The top panel shows the $\Delta$\,C-stat multiplied 
            by the sign of the line in order to distinguish both the significance 
            and the type (emission/absorption-like) of each feature.
            The bottom panel shows the ratio between the normalization of the Gaussian
            and {its error}.
            The top and bottom X-axes show the observed and the rest-frame (z=0.596)
            wavelength range, respectively. The vertical dotted lines indicate the rest-frame transitions of
            the strongest emission lines, whilst the blue dashed ones refer to lines from cool gas. 
            {Shaded grey regions refer to features below the 2\,$\sigma$ detection limit
            and the noise level.}} 
   \label{Fig:Fig_line_search}
\end{figure}

In order to search for any weak lines produced by cool gas and therefore missed by 
our isothermal 7\,keV model we prefer to perform a line scan over the RGS spectrum.
Following the procedure used in \citet{Pinto2016nature}, we search for any line emission
on top of the continuum by fitting a Gaussian scanning over the $6-33$\,{\AA} wavelength range 
with increments of 0.05\,{\AA} and calculating the $\Delta$\,C-statistics.
We adopt a linewidth (FWHM) of 160 km s$^{-1}$, which is the measurement obtained 
from \textit{Hitomi} on Perseus \citep{Hitomi2016nat}. 
We repeat the procedure adopting linewidths 
ranging between 100 and 1000 km s$^{-1}$ without finding major effects on the line detection.
Whilst running the scan we also turn off the line emission from the isothermal gas
in order to check that the lines detected with highest significance indeed correspond
to the strongest lines described by the isothermal model.
The continuum model is basically bremsstrahlung emission.
This is achieved with the {\scriptsize{SPEX}} command \textit{``ions ignore all''}.
We show the results of the RGS line scan in Fig.\,\ref{Fig:Fig_line_search}.
Both the $\Delta$\,C-statistics ($\Delta\,C$) and the ratio between the normalization 
and its uncertainty of the Gaussian are plotted. The $\Delta\,C$ are multiplied by the sign
of the Gaussian normalization to distinguish between emission and absorption-like
features. No absorption line is found at $\Delta\,C$ of 9 or above
corresponding to a normalization/error ratio of 3 or above.

It is straightforward to identify some isolated emission features with the strongest
transitions in this energy band such as {Fe\,\scriptsize{XXIV}} (10.63\,{\AA}), 
{Si\,\scriptsize{XIV}} (6.19\,{\AA}), 
{\nex} (12.135\,{\AA}) and {\oviii} (18.97\,{\AA}).
The {\oviii} $K \alpha$ line is so strong that pops up within a spectral range
that is already affected by the background and it was not shown 
in previous work on Phoenix.
Two other interesting lines are also detected in correspondence 
of the strongest transitions to the ground state of Fe\,{\scriptsize{XXII-XXIII}} and 
{\scriptsize{XXI}} at (11.79\,{\AA}) and (12.34\,{\AA}), respectively,
which indicates that there may be some gas well below 3 keV.
{Additional, fainter, features consistent with weaker Fe\,{\scriptsize{XXI-XXII}} transitions 
are found within 9--10\,{\AA}.}
We do not find strong lines from very cool gas such as Fe\,{\scriptsize{XVII}},
although there is a spike just at 17.1\,{\AA} where the forbidden line
is expected. It is surprising that no Fe\,{\scriptsize{XVII}} resonant line is detected
at 15.0\,{\AA}, but we notice that it falls near the absorption edge of the 
neutral oxygen in the Galactic ISM and where RGS\,2 chip is missing,
yielding half the effective area.
Briefly, the RGS\,2 detector misses the 20--24\,{\AA} observed wavelength range,
while RGS\,1 lost the 10.5--13.5\,{\AA} range. 
Luckily, RGS 1 and 2 complement each other,
but the effective area in the shared bands 
(below 10.5\,{\AA}, 13.5--20\,{\AA} and above 24\,{\AA} in the observed frame) 
is boosted and increases the detectability of some spectral features.
{Another possibility would be that the 15.0\,{\AA} line is highly suppressed by
resonant scattering as suggested for the Perseus cluster in
\citet{Pinto2016mnras}.}

\subsection{Line significance}
\label{sec:line_significance}

The statistical improvement to the best fit of the continuum model yielded by each line
provides a crude estimate of the corresponding line significance. The lines
detected are exactly those expected for a 7\,keV cool core of a massive 
($M\sim 2\times10^{15}\,M_{\odot}$) cluster like Phoenix
and are detected at their rest wavelengths.

In Table\,\ref{table:line_significance} we quote the wavelengths and the statistical improvements of the 
strongest lines detected as expected from their dominant transitions. The Fe\,{\scriptsize{XXII-XXIII}}
lines merge into one feature that is closer to the Fe\,{\scriptsize{XXII}} line centroid. 
The confidence levels (CL) corresponding to the measured $\Delta C$ values are also reported as 
$\%_{MAX}$ and reported as to 99.99 whenever they match or surpass this threshold.

In order to estimate a rigorous and more conservative line significance we need 
to correct for any fake detection and account for the ``look-elsewhere'' effect. 
This is done with Monte Carlo simulations.
We use the bremsstrahlung continuum as a template model to simulate 10,000 spectra with
the same response matrix, exposure time and background as the actual RGS data. 
Then an identical line search is performed as for the original data.
The significance of a line found in the data is computed as the ratio of the number of simulations 
that do not show a line with the same or higher $\Delta C$ value to all performed simulations. 
The results of the simulations are also shown in Table\,\ref{table:line_significance} ($\%_{MCLE}$).
Regardless of the method, several lines are detected well above $3\,\sigma$.
{However, we notice that our lines are not random species detected at random locations. 
Fe\,{\scriptsize{XXI-XXII}}, for instance, are the strongest transitions expected after the well-known 
Fe\,{\scriptsize{XXIII-IV}} lines and are at their expected wavelengths. 
For this reason the Monte Carlo result has to be regarded as a very pessimistic case. 
We examine the Fe\,{\scriptsize{XXII-XXIII}} confusion by including all the lines 
of the best-fit baseline model and check the fluxes predicted by the latest AtomDB and SPEX databases. 
The improvement obtained by adding a further Fe\,{\scriptsize{XXII}} line is $\Delta$\,C-stat\,=\,13 (compared to the total of 15.7), 
which for a single line detection is still significant albeit at lower levels, with Fe\,{\scriptsize{XXII}} providing the 
strongest contribution at 11.79\,{\AA}. The transitions are the strongest expected from AtomDB and SPEX: 
Fe\,{\scriptsize{XXI}} $1s^2 2s^2 2p^1 3d^1 \rightarrow 1s^2 2s^2 2p^2$ (Level $40 \rightarrow 1$)
and Fe\,{\scriptsize{XXII}} $1s^2 2s^2 3d^1 \rightarrow 1s^2 2s^2 2p^1$ (Level $21\rightarrow 1$).}

\begin{table}
\caption{RGS line significance.}  
\label{table:line_significance}      
\renewcommand{\arraystretch}{1.}
 \small\addtolength{\tabcolsep}{-3pt}
\scalebox{0.975}{%

\begin{tabular}{c c c c c c c c c c c c}     
\hline  
         Ion   & {Fe\,\scriptsize{XXIV}} & {Fe\,\scriptsize{XXII-XXIII}}  & {Fe\,\scriptsize{XXI}}   & {Fe\,\scriptsize{XVII}}(f)  \\  
         $\lambda$ (\AA)  & 10.63$\pm$0.01             & 11.79$\pm$0.02         & 12.34$\pm$0.02       & 17.1$\pm$0.02  \\ 
         $\Delta C$           & 40.20                               & 15.70                      & 19.75                     & 9.54                   \\ 
         $\%_{MAX}$       &  99.99                              &  99.99                       &  99.99                    &  99.80                   \\
         $\%_{MCLE}$     &  99.99                             &  99.32                      &  99.95                    &  95.11                   \\
\hline  
         Ion                       & {S\,\scriptsize{XV}}    & {Si\,\scriptsize{XIV}}    & {Ne\,\scriptsize{X}}    & {O\,\scriptsize{VIII}} \\  
         $\lambda$ (\AA)  & 5.06$\pm$0.01         & 6.19$\pm$0.01             & 12.14$\pm$0.01      & 18.97$\pm$0.02      \\ 
         $\Delta C$           & 15.28                         & 35.16                          & 32.3                       & 20.16                     \\ 
         $\%_{MAX}$       &  99.99                        &  99.99                        &  99.99                    &  99.99                   \\
         $\%_{MCLE}$     &  99.46                        &  99.99                       &  99.99                    &  99.95                 \\
\hline                                               
\end{tabular}}

Notes -- Some of the strongest lines detected for each ion. {Fe\,\scriptsize{XXII-XXIII}} lines at 11.7-11.8\,{\AA}
cannot be resolved. Wavelengths account for a redshift of 0.596. $\Delta C$ is the fit improvement
of each line. $\%_{MAX}$ is the maximum confidence level calculated from the $\Delta C$;
 $\%_{MCLE}$ is the confidence level estimated with 10,000 Monte Carlo simulations to account for the 
look-elsewhere effect.
\end{table}


\subsection{Cooling rate}
\label{sec:cooling_rate}

In order to place constraints on the amount of gas cooling from 7\,keV all the way down to 0.1\,keV
we refit the RGS spectrum adding a cooling flow model to the isothermal gas reported 
in Table\,\ref{table:rgs_bestfit}.
The \textit{cf} model in {\scriptsize SPEX} calculates the spectrum of a standard isobaric cooling flow. 
The differential emission measure distribution for the isobaric cooling flow model can be written as
\begin{equation}  \label{Eq:cflow}
D(T) \equiv \frac{dY(T)}{dT} = \frac{5 \dot{M} k}{2 \mu m_{\rm H} \Lambda(T)}
\end{equation}
where $\dot{M}$ is the mass deposition rate, $k$ is Boltzmann's constant, 
$\mu$ the mean molecular weight (0.618 for a plasma with a 0.5 Solar metallicity), 
$m_{\rm H}$ is the mass of a hydrogen atom, and $\Lambda(T)$ is the cooling function. 
The cooling function is calculated by {\scriptsize SPEX} for a grid of temperatures 
and for 0.5 Solar metallicity. The spectrum is evaluated by integrating the above 
differential emission measure distribution between a lower and an upper temperature boundary.

\begin{figure}
  \includegraphics[width=0.985\columnwidth, angle=0]{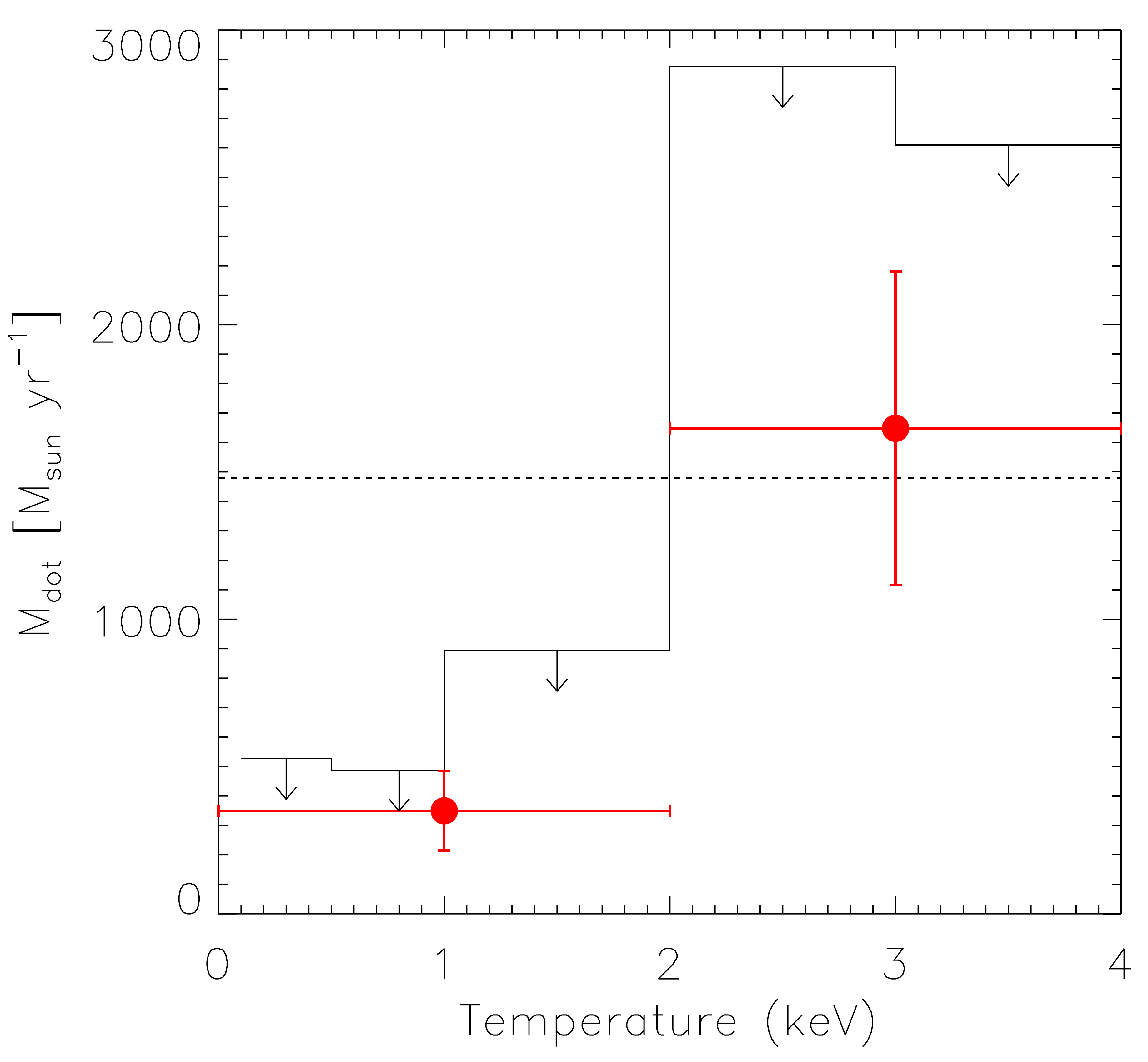}
   \caption{Constraints on cooling rates using two classical radiative cooling flow 
   models (red points) and upper limits (black arrows) for a five-component cooling model on top of the 
   isothermal {\scriptsize{CIE}} model.} \label{Fig:Fig_IDL_cooling_rates}

\end{figure}

At first we use a complex model with five \textit{cf} components plus the isothermal \textit{cie}
all corrected by redshift and absorption. The aim is to understand how much gas is cooling
between several temperature ranges (0.1--0.5, 0.5--1.0, 1.0--2.0, 2.0--3.0, 3.0--7.0 keV).
The only free parameter in each \textit{cf} is the cooling rate, $\dot{M}$, whilst the abundances
are coupled to those of the \textit{cie} component, which are constrained by the strongest lines.
Obviously, there is a high degeneracy in the measurements due to low statistics owing to 
the large distance of the Phoenix cluster. We can only measure upper-limits on the amount
of gas cooling within each temperature interval (see Fig.\,\ref{Fig:Fig_IDL_cooling_rates}). 
Despite the degeneracy, a significant drop of gas cooling below 2\,keV might be present.

In order to place tighter constraints and not to boost the degeneracy, we simplify the model
by using only two cooling flow components (0.1--2.0, 2.0--7.0 keV) in addition of the isothermal 
gas model.
We estimate that $350\pm_{120}^{150} (68\%) \pm_{200}^{250} (90\%)\,M_{\odot}\,{\rm yr}^{-1}$ 
are cooling below 2\,keV.
Obviously, this simple model broadly agrees with the more complex five \textit{cf} 
components (see Fig.\,\ref{Fig:Fig_IDL_cooling_rates}).

We also refit the RGS spectrum with a simple two-temperature model consisting of the baseline \textit{cie} 
model plus an additional \textit{cie} component. This additional cool component improves the fit
by $\Delta C = 20.5$ for two additional degrees of freedom 
($n_{\rm e} n_{\rm H} V_2 = 8.5\pm2.5 \times 10^{72} {\rm m}^{-3})$ and $kT_2 = 1.20\pm0.13$ keV,
mainly constrained by the Fe\,{\scriptsize{XXI-XXII}} lines).
{Excluding these lines the fit improvement would be much smaller ($\Delta C = 7.5$) with 
$T \sim$ 1.1 keV.}

%

\subsection{Instrumental broadening}
\label{sec:instrumental_broadening}

The spectral fits yield a maximum line width corresponding to a velocity broadening 
$v_{\rm mic} (1D) < 420 $ km s$^{-1}$ (90\% upper limit or $\Delta C = 2.71$). 
This measurement, however, does
not account for the fact that some of the broadening is due to the nature of the RGS detectors.
The instrumental broadening can be subtracted using the surface brightness computed with
the CCD detectors. This is achieved through the use of the \textit{lpro} model in {\scriptsize{SPEX}}.
The \textit{lpro} receives as input the spatial broadening measured with the MOS\,1 surface brightness
profile and convolves it with the baseline spectral model:

Spectrum = ( \textit{hot} $\times$ (redshift $\times$ \textit{cie}) ) $\circledast$ \textit{lpro}.

The \textit{lpro} component now accounts for the spatial broadening and the $v_{\rm mic}$ parameter 
of the \textit{cie} component accounts for any leftover line broadening due to turbulence.
A new fit of the RGS spectra with this model place the 90\% upper limit of the velocity broadening
down to 280 km s$^{-1}$.

\citet{Sanders2013} have shown that the CCD spatial profile is highly affected
by the large scale distribution of the hot gas which mostly contributes to bremsstrahlung 
continuum rather than lines. The physical scale of the hot gas is much larger than that of the cool core 
that produces the emission lines detected in the RGS spectra (see, e.g., \citealt{Pinto2016mnras} 
for more detail on the link between the line widths and the location of the cool gas). 
This means that using the full spatial profile might result into an over-subtraction of the 
spatial broadening and, therefore, an under-estimate of the upper limit on the turbulence.
\citet{Sanders2013} attempted at subtracting spatial profiles extracted in energy bands that include
only strong lines. 
They obtained typical levels of 300--400 km s$^{-1}$ with systematic uncertainties 
of about 150 km s$^{-1}$.

In \citet{Bambic2018} we introduced a new method which consists of fitting three Gaussian lines to the 
spatial profile. The central Gaussian can be interpreted as the coldest core of gas, the outer Gaussian 
reproduces the bremsstrahlung continuum from hot hydrogen gas beyond the cluster core. 
The intermediate Gaussian models any smooth transition between cool and hot regions.
This may prove to be necessary for highly multi-phase and nearby clusters such as Centaurus,
but not crucial for an object like Phoenix. 
The narrowest Gaussian obtained by fitting a single Gaussian, two Gaussian and three Gaussian lines
to the MOS spatial profiles are shown in Fig.\,\ref{Fig:Phoenix_MOS_profile}.
We test each of these three profiles independently as input of the \textit{lpro} model as
shown above. We obtain $v_{\rm mic} (1D)$ 90\% upper limits of about 400, 370 and 330 km s$^{-1}$
using the narrowest of one, two and three Gaussian lines, respectively.

{We also extract a spectrum in a smaller region (70\% of the PSF, i.e. 100 kpc)
and, interestingly, find that the line width strongly decreases with a 2\,$\sigma$ (90\%) 
upper limit of only 290 km s$^{-1}$  (without any spatial broadening subtraction). 
This further supports our results obtained with the 90\% PSF 
through the modelling of the spatial broadening.}

{Finally, we extract \textit{Chandra} ACIS surface brightness profiles from the longest exposures 
along the RGS dispersion direction and produce spatial broadening profiles similarly to EPIC/MOS 1. 
We then fit models with 1, 2 and 3 Gaussians to measure the core spatial broadening. 
We find that the limits obtained using the \textit{Chandra} data are systematically larger than those obtained 
with MOS data, but by only a small amount (10-15 km s$^{-1}$).}

\subsection{Turbulence required for heat transfer}
\label{sec:heat_transfer}

In \citet{Bambic2018} we showed that the balance between cooling and heating at any radius $r$ 
within the cool core requires a minimum level of turbulence. This is in order to propagate the energy 
from the central AGN to that location. This translates into the following threshold:
\begin{equation} \label{Eq:vel}
\sigma_{1D ({\rm km/s})} = 4.69 \times \bigg(  \frac{r_{\rm kpc} \, L_{44}}{M_{15}} \bigg)^{1/3}
\end{equation}
where $L_{\rm cool} = L_{44} \times 10^{44} {\rm ergs/s}, 
M_{\rm gas} = M_{15} \times 10^{15} M_{\odot}, r = r_{\rm{kpc}} \times 3.09 \times 10^{21} \rm{cm}$.
In \citet{Bambic2018} we noticed that this approach provides conservative limits 
for the required propagation velocity of turbulence.

\begin{figure}
  \includegraphics[width=0.99\columnwidth, angle=0]{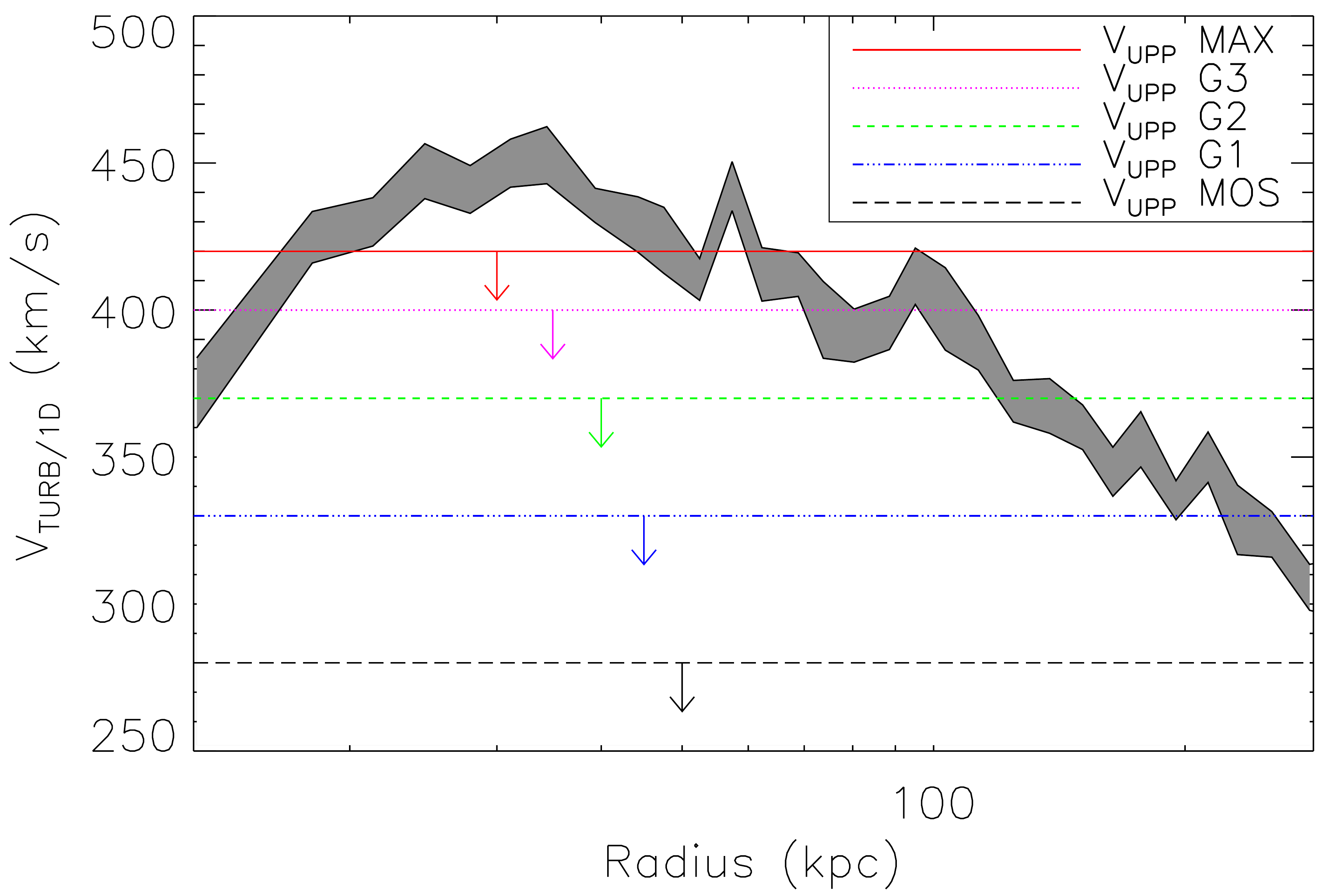}
   \caption{Constraints on turbulence. The grey region shows the turbulence required to 
   propagate the heat throughout the cluster from the central AGN up to 250 kpc. 
   The lines show the 90\% upper limits obtained with the RGS spectral fits: total line width
   (top dotted line) compared to leftover broadening after subtracting the narrow Gaussian fits 
   (dashed/dotted-dashed lines) or the full spatial MOS profile (bottom dashed line).} 
   \label{Fig:Fig_IDL_VMIC_constraints}
\end{figure}

We use \textit{MBProj2} \citep{Sanders2018} to produce 
profiles for $L_{44}$ and $M_{15}$ from \textit{Chandra} observations. 
The gas mass is estimated from the observed surface brightness with the assumptions 
of spherical symmetry and hydrostatic equilibrium. 
The profiles are computed assuming an NFW dark matter mass profile, 
using the same data, analysis procedure and BIN-NFW model as described in \citet{Sanders2018}.
The 68\% statistical uncertainties on the turbulent velocity are computed from Monte Carlo fits of 
Eq.\,\ref{Eq:vel} to the velocity profiles, implementing error chains on each of the measured quantities.
The error bars on the turbulent velocity calculated at each radius are shown in 
Fig.\,\ref{Fig:Fig_IDL_VMIC_constraints} and are compared with upper limits on turbulence
as measured with the linewidths in the RGS spectra. 
It is clear that regardless of how severe is the subtraction of the instrumental broadening,
the best constrained 90\% upper limits on $v_{\rm mic} (1D)$ lie below the velocity required 
to transfer heat throughout the cool core and offset radiative cooling,
particularly within the inner 100 kpc.
{The use of a narrower extraction region and the spatial broadening profiles computed with 
\textit{Chandra} supports these results.}

\section{Discussion}
\label{sec:discussion}
 
\subsection{Cooling and star formation rates}
\label{sec:discussion_cooling}
 
In this work we study the Phoenix cluster (SPT-CLJ2344-4243) in order to understand
why such a massive cluster exhibits a strong starburst of $500-800\,M_{\odot}\,{\rm yr}^{-1}$
in combination with a powerful AGN and a very low central cooling time (see, e.g., \citealt{McDonald2012}).
This object is of high interest as it is also the most luminous X-ray cluster known 
and the most distant object in our XMM-\textit{Newton}/RGS legacy catalog (Pinto et al. in prep).
Phoenix is currently the only cluster at redshift 0.6 for which high-resolution X-ray spectroscopy
is feasible. This technique is crucial in order to obtain reliable constraints on
turbulence, abundances and emission measures of cool gas phases. 
Phoenix can also be regarded as an early stage and proxy
for nearby, massive, cool core clusters.

We find that the amount of cool gas below 2 keV is consistent with a cooling rate of 
$350\pm_{120}^{150} (68\%) \pm_{200}^{250} (90\%)\,M_{\odot}\,{\rm yr}^{-1}$ (see Fig.\,\ref{Fig:Fig_IDL_cooling_rates}), 
which agrees with the upper limits obtained by \citet{Tozzi2015}.
We detect for the first time evidence of {O\,\scriptsize{VIII}} and {Fe\,\scriptsize{XXI-XXII}}
emission lines in the Phoenix cluster. These iron lines demonstrate the presence of gas below 2 keV.
Our measured cooling rate is broadly consistent with the starburst rate,
suggesting a possible link between the gas cooling in the inner core ($<100$ kpc)
and that involved in the starburst. 
{Interestingly, the spectrum extracted in a narrower region (70\% of the PSF, 
see Sect.\,\ref{sec:instrumental_broadening}) 
shows that the overall flux of the hot component decreases by 13\%, while the cold component 
is consistent with being constant. This suggests that the cold gas is located in the innermost regions.}

The {tentative detection} of {Fe\,\scriptsize{XVII}} forbidden 
and intercombination lines and the absence
of the corresponding resonant line would suggest that this cool phase might be subject to
some form of heating. This would agree with the detection of strong {O\,\scriptsize{VI}}
in HST/COS spectra of the Phoenix cluster (\citealt{McDonald2015}).
Unfortunately, the {O\,\scriptsize{VII}} lines are hidden within the noise
and the high instrumental background around 35\,{\AA}; here the spectrum
does not provide reliable constraints. 
However, we successfully model the {Fe\,\scriptsize{XVII}} 
emission lines and other cool gas features with a photoionized gas model.
This is done by adding into our {\scriptsize{SPEX}} best-fit model 
an additional emission component through the {\scriptsize{PHOTEMIS}}
code\footnote{https://heasarc.gsfc.nasa.gov/xstar/docs/html/node106.html} 
adopting a low ionization parameter of about 1 erg s$^{-1}$ cm.
Our model predicts a total {O\,\scriptsize{VI}} luminosity of $10^{43}$ erg s$^{-1}$ or above, 
in agreement with the accurate measurements obtained
in the FUV with HST/COS spectra (\citealt{McDonald2015}).

The picture might be slightly more complicated since the molecular and neutral gas in the filaments could absorb
some of the soft X-ray emission. This means that the cooling rate that we measure is possibly a lower limit.
Unfortunately, without knowing the covering fraction and solid angle of this coolest phase 
as well as its location with respect to the soft X-ray emitting gas, it is difficult to
estimate its column density and thus the unabsorbed soft X-ray emission.

\subsection{Cooling versus heating propagation}
\label{sec:discussion_heating}
 
The presence of both high star formation and cooling rates in the Phoenix cluster 
casts doubts on the efficiency of AGN feedback to balance cooling in this system. 
However, the AGN is certainly affecting the cluster and this can be seen, for instance, 
in the ALMA observations of the CO(3-2) line emission which show that approximately 
half of the molecular gas, $\sim1\times10^{10}\,M_{\odot}$, is located mainly at the peripheries of the radio bubbles (\citealt{Russell2017}).

It is interesting to note that our cooling rate of 
$350\pm_{120}^{150} (68\%) \pm_{200}^{250} (90\%)\,M_{\odot}\,{\rm yr}^{-1}$
is high enough to produce the $160\pm50\,M_{\odot}\,{\rm yr}^{-1}$ necessary to 
cool down into the molecules that are found in the filaments ($1\times10^{10}\,M_{\odot}$), 
assuming a buoyant rise time of 50--120 Myr for the bubbles
(e.g., \citealt{Russell2014}, \citealt{McNamara2014}, \citealt{Russell2017}). 
Moreover, if we decouple the redshifts of the two thermal components
in our RGS two component spectral fits (see bottom of Sect.\,\ref{sec:cooling_rate}), 
we find that the cool gas (1.2\,keV, {Fe\,\scriptsize{XXI-XXII}}) has a systematic velocity of 
$+1000\pm400$ km s$^{-1}$ on top of the redshift of the main 7\,keV component. 
Despite the large uncertainties, this result resembles the velocity shift of 250 km s$^{-1}$
found in the molecular gas, suggesting that these two phases may actually be related 
and that they are falling back towards the nucleus of the brightest cluster galaxy (BCG).

In order to place tight constraints on turbulence and estimate whether the energy stored
in turbulent motions is high enough to balance radiative cooling,
we perform an accurate high-resolution spectroscopic analysis. 
The total line width of the emission lines detected in the RGS spectrum is below
420 km s$^{-1}$ (90\% upper limit of the 1D micro-turbulence velocity). 
Using a newly introduced technique which accounts for instrumental broadening \citep{Bambic2018},
we show that the turbulence level in the Phoenix cluster is likely well below 400 km s$^{-1}$.
{Spectra extracted in narrower regions yield even lower limits (290 km s$^{-1}$,
see Sect.\,\ref{sec:instrumental_broadening}).}
This result suggests similar values of turbulence detected in the Perseus cluster with 
\textit{Hitomi} on similar physical scales (\citealt{Hitomi2016nat}), in the core of nearby giant 
elliptical galaxies through resonant scattering techniques (\citealt{Ogorzalek2017}) 
and several nearby clusters (\citealt{Sanders2013}, \citealt{Pinto2015}).

Therefore, turbulent motions alone are below the level required
to balance radiative cooling loss (see Fig.\,\ref{Fig:Fig_IDL_VMIC_constraints}).
This means that dissipation of turbulence by itself may not provide 
and propagate the amount of energy
required to quench cooling (and star formation) in the cool core
and gives a natural explanation for the coexistence of the powerful AGN, 
strong cooling and starburst.
AGN feedback and its effects may be insufficient
to compete with the massive amount of cool gas in the core of the cluster.
Alternatively, it is possible that the energy necessary to quench
cooling is transported via sound waves as suggested 
by \citet{Fabian2017sw} and \citet{Bambic2018}.
The high star formation itself may be provoked by 
Compton cooling due to the strong AGN radiation 
(e.g. \citealt{McDonald2015} and \citealt{Walker2014}).
{In the outer regions, weak shocks produced by mergers and sloshing, 
travelling at mildly supersonic velocities, may provide a further source of heating
(e.g. \citealt{Ascasibar2006}, \citealt{Lau2009} and Walker et al. submitted)}.

The abundance pattern that we measure in the Phoenix RGS spectrum is consistent with the 
estimates obtained with CCD spectra \citep{Tozzi2015} and broadly agrees
with the abundances commonly found in nearby cool core clusters (\citealt{Mernier2016}).
This suggests that there are not significant differences between the Phoenix cluster,
a prototype high-$z$ massive cluster, and the nearby cool core clusters. 

It is striking to notice that the Phoenix cluster exhibits high cooling and star formation
rates if compared to smaller cool cores where the predicted cooling rates are much larger
than than the actual measurements of cooling rates and star formation rates. 
According to \citet{McDonald2018} there is a significant 
trend in the ratio between the star formation and the cooling rates above 
$\sim30\,M_{\odot}\,{\rm yr}^{-1}$.
They interpret the trend as an increasing efficiency with which the ICM cools and 
forms stars in the strongest cool cores such as Abell\,1835 
($\sim140\,M_{\odot}\,{\rm yr}^{-1}$, see \citealt{Sanders2010}) 
and Phoenix ($\sim350\,M_{\odot}\,{\rm yr}^{-1}$, our value).
As \citet{McDonald2018} pointed out, large cooling rates may result in high
accretion onto the supermassive black holes, which can progressively switch
from a kinetic/radio to a quasar/radiative feedback mode.
This could somehow lower the efficiency of quenching star formation
due to a possible lower energy release by radiation 
compared to powerful radio jets (see e.g. \citealt{Churazov2005})
and change the way cooling actually occurs with lower ICM emission
of soft X-rays (Compton cooling, see also \citealt{Walker2014}).

\subsection{Future missions: \textit{XRISM} and \textit{ATHENA}}
\label{sec:discussion_missions}
 
Current X-ray instruments that enable direct velocity measurements such as the gratings
on board XMM-\textit{Newton} and \textit{Chandra} and their CCD/imaging detectors that
allow us to obtain indirect velocity measurements have several limitations 
(see, e.g., Sect.\,\ref{sec:intro} and references therein). 

The most accurate, direct, measurement of line broadening and, therefore,
constraint on turbulence, was obtained by the {\textit{Hitomi}} satellite
during its deep (230ks) observation of the Perseus cluster of galaxies \citep{Hitomi2016nat}. 
The soft X-ray spectrometer (SXS) microcalorimeter onboard {\textit{Hitomi}}
enabled to measure an average line broadening of $164\pm10$\,km\,s$^{-1}$
and possible lower than 100 km\,s$^{-1}$ once PSF effects were taken into account
\citep{Hitomi2017atm}. This started to cast doubts on the efficiency of turbulence to propagate the heat
throughout the cluster rapidly enough to offset radiative losses 
(see also \citealt{Fabian2017sw}).

Due to a chain of unfortunate circumstances, {\textit{Hitomi}}
(also known as \textit{ASTRO-H}, \citealt{Takahashi2010}) was lost,
but a new X-Ray Imaging and Spectroscopy Mission (\textit{XRISM},
a.k.a \textit{XARM}\,\footnote{https://heasarc.gsfc.nasa.gov/docs/xarm/}, 
X-ray Astronomy Recovery Mission)
has been accepted and will be likely launched in 2022
(see, e.g., \citealt{Guainazzi2018}).

\begin{figure}
  \includegraphics[width=0.975\columnwidth, angle=0]{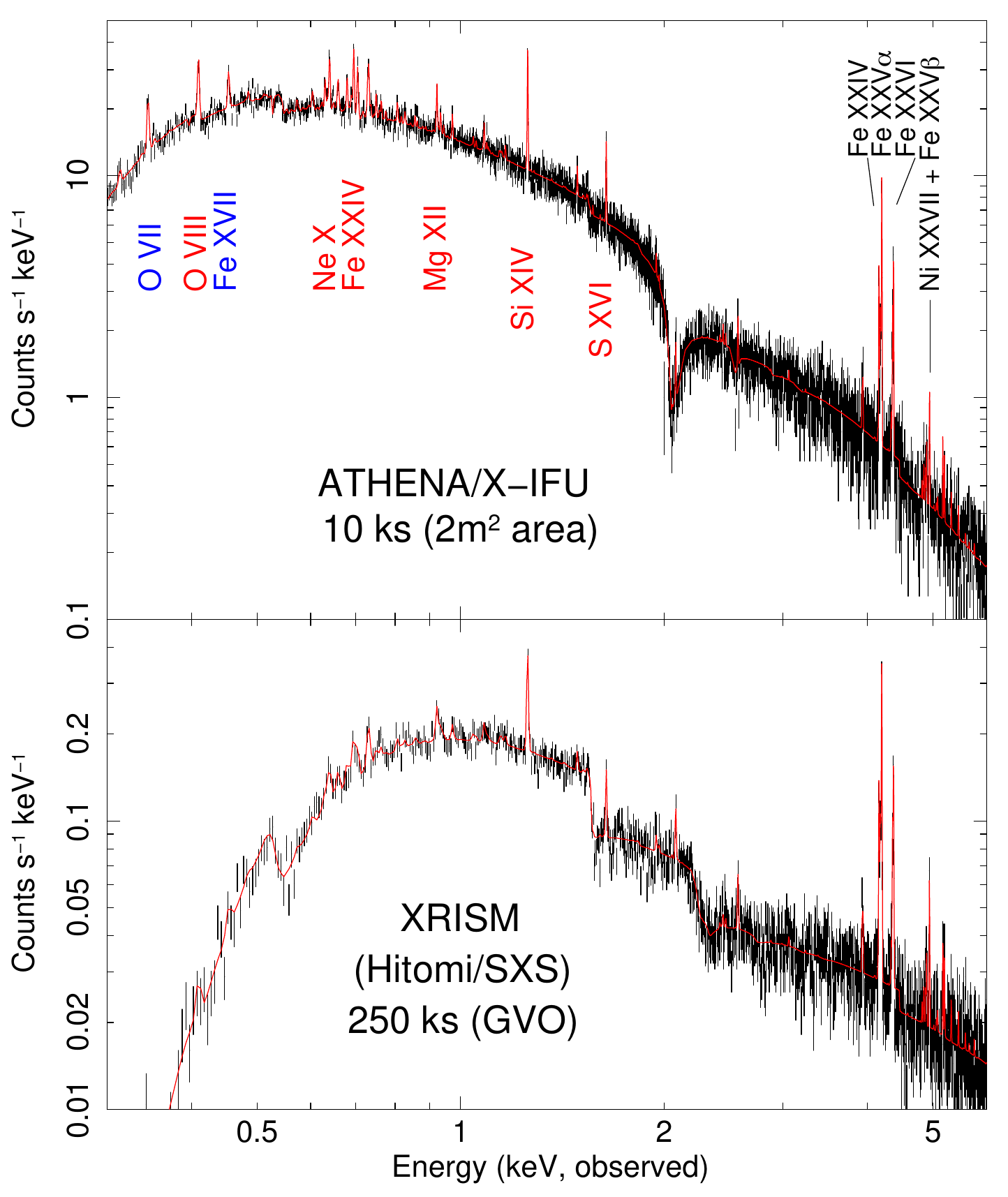}
   \caption{Simulations of the Phoenix cluster for a region of 50 arc second size
                 adopting the XMM-\textit{Newton}/RGS best fit model
                 plus the AGN and the cool ($<2$ keV) gas (see text).
                 Cosmic X-ray background and particle background are accounted for.
                 The strongest transitions are labelled in the observed frame (we adopt $z=0.596$
                 and turbulence broadening $v_{1D} = 250$ km s$^{-1}$).
                 We use the response matrix of {\textit{Hitomi}}/SXS as a proxy for {\textit{XRISM}}.} 
   \label{Fig:Fig_Phoenix_XIFU_SXS_sim}
\end{figure}

Here we perform a simulation of the Phoenix cluster as it will be seen by 
{\textit{XRISM}} using the {\textit{Hitomi}}/SXS 5\,eV response matrix 
(see Fig.\,\ref{Fig:Fig_Phoenix_XIFU_SXS_sim}, bottom panel).
We adopt our XMM-\textit{Newton}/RGS best fit model 
consisting of two thermal models
with temperatures of 1.2 and 7.0 keV, see Sect.\,\ref{sec:cooling_rate} 
and Table\,\ref{table:rgs_bestfit} plus photoionized gas that reproduces the 
{Fe\,\scriptsize{XVII}} X-ray emission and the {O\,\scriptsize{VI}} UV flux as measured 
by \citet{McDonald2015}. We assume 250ks exposure time, similar to 
the Perseus cluster, but with the gate valve open (GVO).
We also take into account the emission from the central AGN by adding
the model of \citet{Tozzi2015}, which consists of an intrinsically-absorbed 
($N_{\rm H} = 5 \times 10^{23}$ cm$^{-2}$) power-law with slope 1.8 
and an unabsorbed intrinsic luminosity of $5 \times 10^{45}$ erg s$^{-1}$
in the 2-10 keV energy band.
We remind the reader that the RGS template model is obtained for a region that
can be approximated with a square of 50 arc second side.

We adopt a line broadening of 250 km s$^{-1}$, which is well within
our upper-limits obtained by subtracting the spatial broadening. 
Despite the much larger distance of the Phoenix ($z=0.6$) compared
to the Perseus cluster ($z=0.018$), we will obtain a statistical uncertainty 
of just 25 km s$^{-1}$ (i.e. 10\%) for a 250ks {\textit{XRISM}} observation.
This is due to the brightness of the Phoenix cluster and its high temperature,
which boosts the Fe K ({Fe\,\scriptsize{XXIV-XXVI}}) emission lines
where the {\textit{XRISM}} resolving power peaks
(see Fig.\,\ref{Fig:Fig_Phoenix_XIFU_SXS_sim}, bottom panel).
This accuracy is high enough to distinguish between some different heating mechanisms.

The X-ray Integral Field Unit (X-IFU) on board the 
Advanced Telescope for High Energy Astrophysics ({\textit{ATHENA}},
\citealt{Nandra2013}) will provide a drastic improvement in both spatial
resolution (from 1 arc minute down to around 5 arc seconds), spectral
resolution (from 5 eV to 2.5 eV) and effective area (more than an order of
magnitude larger than any previous grating or microcalorimeter).
For the astrophysics of galaxy clusters this means an improvement of
two orders of magnitude in our capabilities.
In Fig.\,\ref{Fig:Fig_Phoenix_XIFU_SXS_sim}, top panel, we show
a simulation of the Phoenix cluster made with the same template model used
above, but with a short exposure time of just 10ks.
Both the foreground emission from our Galaxy, the Local Bubble 
and non X-ray background due to highly-energetic particles are
taken into account following the mission requirements\footnote{http://www.the-athena-x-ray-observatory.eu/resources/simulation-tools.html}.
The sky region is a square of 50 arc second side as above.

The X-IFU 10ks snapshot of the Phoenix cluster will provide a spectacular spectrum 
of high quality and yield statistical uncertainties of 9 km s$^{-1}$ (i.e. $\sim4$\%)
on the velocity broadening.
If we include - in quadrature - additional systematic uncertainties due to gain homogeneity, 
stacking of nearby pixels within the 50 arc second area and uncertainty in the spectral resolution,
we may get an uncertainty of up to 15-20 km s$^{-1}$ (i.e. $6-8$\%,
which is well above the 5\,$\sigma$ level).
The current mission requirement for the effective area is
1.4 m$^2$ at 1 keV. This will not produce any major issues for these measurements
because the accuracy is driven by the high-energy band above 2 keV
apart from a small increase of the exposure time.
Moreover, the 5 arc second spatial resolution of the X-IFU will enable us
to perform accurate measurements of turbulent velocities 
in more than 20 different regions of the cool core
for an exposure time of just 100 ks.

The Phoenix cluster is an extraordinary, bright, cooling core cluster producing X-ray spectra
with strong emission lines. However, the X-IFU will enable accurate turbulence measurements
with $10-20$\% uncertainties (i.e. 5\,$\sigma$) for several clusters at high redshift 
 - at high AGN feedback and star formation rates - and fairly short (50-100 ks) exposures.

\section{Conclusions}
\label{sec:conclusion}

The Phoenix cluster (SPT-CLJ2344-4243) is
the most luminous X-ray cluster known and the most distant cluster observed 
with high-resolution X-ray spectrometers.
It exhibits an outstanding starburst of $500-800\,M_{\odot}\,{\rm yr}^{-1}$,
which is among the largest found in the local Universe ($z<1$),
and an AGN inflating bubbles via powerful radio jets.
The presence of high star formation and cooling rates in combination 
with a powerful AGN is puzzling if compared to the low rates shown by nearby 
cooling core clusters. 
In this work, we perform an in-depth study of the XMM-\textit{Newton}
observations of the Phoenix cluster with particular focus on the high-resolution
spectrometers in order to obtain accurate measurements of the cooling rate
and turbulent broadening.

We find that about $220-480\,M_{\odot}\,{\rm yr}^{-1}$ are cooling below 2\,keV 
in the cluster core, in broad agreement with the estimates of star formation rate, 
and that the turbulence level is likely not adequate 
to fully propagate AGN heating throughout the cool core.
This provides a natural explanation for the coexistence of a large amounts of cool gas,
star formation and a powerful AGN in the Phoenix cluster.
However, the comparison with nearby, more quiescent, cool core clusters
suggests that the discrepancy is likely due to either to the younger age
of the Phoenix cluster, and therefore of the BCG AGN feedback activity, or 
the accretion rate on the supermassive black hole and the feedback mode.

Future missions such as \textit{ATHENA} and \textit{XRISM} will boost the 
accuracy in measuring crucial parameters in X-ray spectra of clusters 
of galaxies. In particular, we show that for Phoenix-like clusters it is possible to 
obtain excellent results - such as a 5\,$\sigma$ detection of turbulence - 
with short snapshots of about 10ks exposure.


\section*{Acknowledgments}

This work is based on observations obtained with XMM-\textit{Newton}, an
ESA science mission funded by ESA Member States and USA (NASA).
We acknowledge support from ERC Advanced Grant Feedback 340442.
We also thank J. de Plaa and J. Kaastra for support in optimising SPEX.
{We also acknowledge the anonymous referee for useful comments 
that have improved the clarity of the manuscript.}


\bibliographystyle{mn2e}
\bibliography{bibliografia} 

\bsp

\appendix

\section{RGS background}
\label{sec:appendix}

\begin{figure}
  \includegraphics[width=0.975\columnwidth, angle=0]{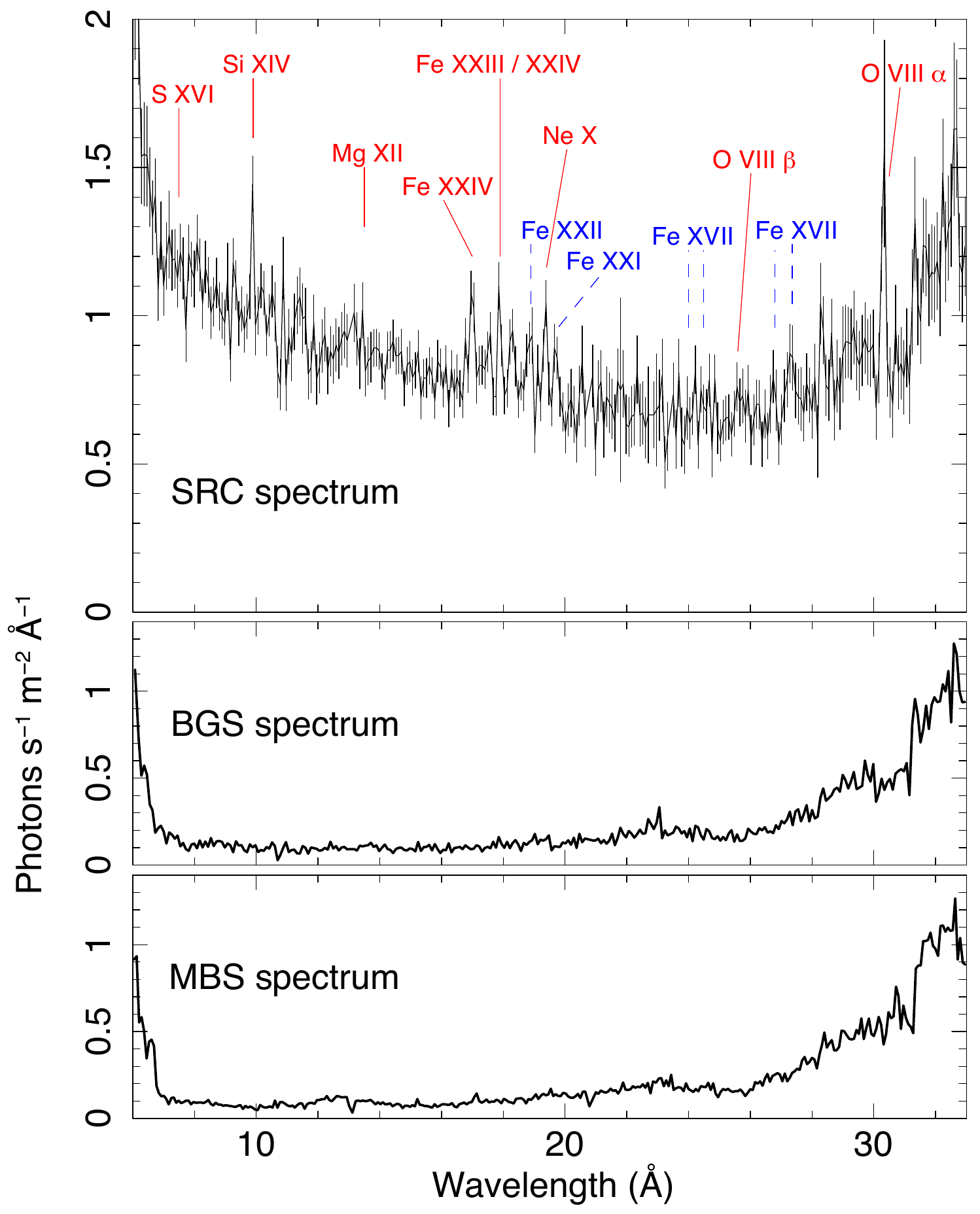}
   \caption{XMM-\textit{Newton}/RGS raw spectra for the source (top panel,
                 without any background subtraction), the background as determined 
                 from the exposure outside the 98\% of the PSF (BGS, middle panel)
                 and the model background computed from templates from blank field 
                 observations (MBS, bottom panel).} 
   \label{Fig:Phoenix_raw_spectra}
\end{figure}

In this section we briefly compare the raw source spectrum of the Phoenix cluster
and two background spectra computed with different methods. 
As mentioned in Section\,\ref{sec:data_spectra},
the RGS 1 and 2 spectra are extracted using as centroid 
$(\alpha, \delta)=(23:44:43.9,-42:43:13.7)$ for all observations and a width of 50 arcsec 
(\textit{xpsfincl} $=90$ in the \textit{rgsproc}).
In order to subtract the background we test both the standard background
spectrum which is extracted beyond the 98\% of the RGS point spread function 
(PSF, \textit{xpsfexcl} $=98$ in the \textit{rgsproc}) and the model background spectrum, 
which is a template background spectrum computed using blank field observations and
normalized by the count rate in CCD\,9. Both are computed by the \textit{rgsproc} task.
Extended, nearby ($z<0.2$), clusters would contaminate the background 
extraction region ($<2'$) with source photons.
However, the Phoenix cluster has a high distance ($z=0.6$) and the background spectrum 
extracted beyond the 98\% of the PSF is rather clean and is comparable to the model 
background spectrum. For display purposes we stack the source raw spectra,
the observation and model background spectra and show them in 
Fig.\,\ref{Fig:Phoenix_raw_spectra}.
We notice that the background spectra are similar and, therefore, the choice of using 
the one or the other does not affect the results. 
The source is brighter than the background up to about 30\,{\AA}, 
where the {\oviii} line clearly exceeds it.

\label{lastpage}

\end{document}